# The Truth about Power Laws: Theory and Reality

Xiaojun Zhang[1*], Zheng He[2*], Liwei Zhang[3], Lez Rayman-Bacchus[4], Yue Xiao[1], Shuhui Shen[1]

[1]School of Mathematical Sciences, [2]School of Management and Economics, University of Electronic Science and Technology of China, Chengdu, P. R. China

[3] School of Mathematical Sciences, Dalian University of Technology, Dalian, P. R. China

[4] Business School, University of Winchester, Winchester, UK

[*]Corresponding author. E-mail: sczhxj@uestc.edu.cn; hezh@uestc.edu.cn

**Consensus about the universality of the power law feature in complex networks is experiencing profound challenges. To shine fresh light on this controversy, we propose a generic theoretical framework in order to examine the power law property. First, we study a class of birth-and-death networks that is ubiquitous in the real world, and calculate its degree distributions. Our results show that the tails of its degree distributions exhibits a distinct power law feature, providing robust theoretical support for the ubiquity of the power law feature. Second, we suggest that in the real world two important factors, network size and node disappearance probability, point to the existence of the power law feature in the observed networks. As network size reduces, or as the probability of node disappearance increases, then the power law feature becomes increasingly difficult to observe. Finally, we suggest that an effective way of detecting the power law property is to observe the asymptotic (limiting) behaviour of the degree distribution within its effective intervals.**

In 1999, Barabási and Albert published a seminal article 'Emergence of scaling in random networks' in *Science* (*1*), in which they suggest that growth and preferential attachment are two key characteristics for real world networks, and further suggest that networks with the power law feature widely exist in the real world. Over the last 20 years, their contribution has exerted a significant impact on network science research (*2-5*).

Generally, a network with the power law feature means that the tail of its degree distribution follows a power law (*6-11*). In other words, there is a positive integer $k'$, and when $k > k'$, the probability of a node with degree $k$ is

$$P(k) \propto k^{\alpha}$$

where $\alpha$ is the exponent of the power law.

For a long time now, researchers and practitioners have believed that the power law feature is common in real world networks, such as the Internet, scientist co-authoring networks, metabolic networks, and biological networks (*12-21*), and numerous results are based on this property (with more than 35,000 citations by *Google Scholar*). However, more recently some have begun to question its universality (*11, 22-26*). For example, Tanaka (*22*) argues that the metabolite degree distributions at the module level follow an exponential distribution. Similarly, Lima-Mendez and Helden (*25*) find that the degree distributions in many biological networks are not subject to a power law. More broadly, Stumpf and Porter (*11*) argue that 'most reported power laws lack statistical support and mechanistic backing'. In addition, Broido and Clauset (*26*), employing statistical tools, analyzed nearly 1000 networks in social, biological, technological, and informational domains, and concluded that scale-free networks are empirically rare. These contradictory and critical studies have shaken the cornerstone of complex network theory, provoking widespread controversy.

Such empirical challenges have spurred researchers to employ alternative or additional criteria to describe the degree distribution, such as low degree saturation, high degree cutoffs, and improved goodness of fit (*7,8,21,27,28*). These refinements are thought to provide a better understanding of deviation from the pure power law. Still, some continue to doubt the reliability or validity of empirical data (*9,29,30*).

Clearly real world networks are dynamic, a feature that must be captured in any attempt to explain power law behavior. We suggest a critical method for informing this debate is to first establish a generic, theoretical, and realistic evolutionary mechanism (*11,21,26*) in order to examine the power law feature. This theoretical mechanism includes four steps: theoretical model building, degree distribution solving, power law feature judgment, and empirical results interpretation.

***Do complex networks with power law feature widely exist in the real world?***

For any network, there are two basic elements: nodes and edges connecting different nodes. In numerous real world networks (e.g. social, ecological, business, and biological), nodes are agents with life-cycles, and may possess intelligence. During the evolving processes of these networks, nodes may enter or exit randomly, reflecting the birth and death of network nodes (*31-34*). Meanwhile, these agents (nodes) may exhibit differing capabilities, making them unequal in terms of resources and positions in their networks. Those with scarce resources or occupying critical positions will be more attractive, and new entrants will prefer to establish linkages with nodes that provide what they need. These behaviors make preferential attachment (*35-39*) widespread in real world networks like the "rich-get-richer" phenomenon (*40-43*). Accordingly, we suggest that random addition and deletion of nodes and preferential attachment are two universal behaviours in the evolution of real world networks.

Based on the above, we establish a network model characterized by random addition and deletion of nodes as well as preferential attachment. The evolving rules are as follows:

i. the initial network is a complete graph with $m+1\left(m\geq 1\right)$ nodes;

ii. at each unit of time, randomly delete a node from the network with probability $q\left(0\leq q<1/2\right)$; or add a new node to the network with probability $p=1-q$ and connect it with $m$ old nodes of network by preferential connection. That is, the probability that the new node connects with an old node $i$ depends on the degree $k_i$ of node $i$, i.e. $\pi_i=\frac{mk_i}{\sum_j k_j}$. (Appendix I).

In the case of $q=0$, this model equates to the BA model. Thus, the BA model is a special case of our model.

Employing the stochastic process rules (SPR) method (*44*), we obtain the equation of (steady-state) degree distribution $K$ for differing $k$ (Appendix II for details).

$$\left(p+\frac{k}{2}+kq\right)P\left(k\right)=\begin{cases} qP\left(1\right) & k=0 \\ \left(m+1\right)qP\left(m+1\right)+\frac{m-1}{2}P\left(m-1\right)+p & k=m \\ \left(k+1\right)qP\left(k+1\right)+\frac{k-1}{2}P\left(k-1\right) & 1\leq k, k\neq m \end{cases}$$

where $P\left(k\right)=P\left\{K=k\right\}$.

By using the probability generating function, the solutions to the above equations are obtained (Appendix III for details). In order to closely observe the property of its tails, Fig. 1 illustrates the solution for $m = 4$ with different $q$.

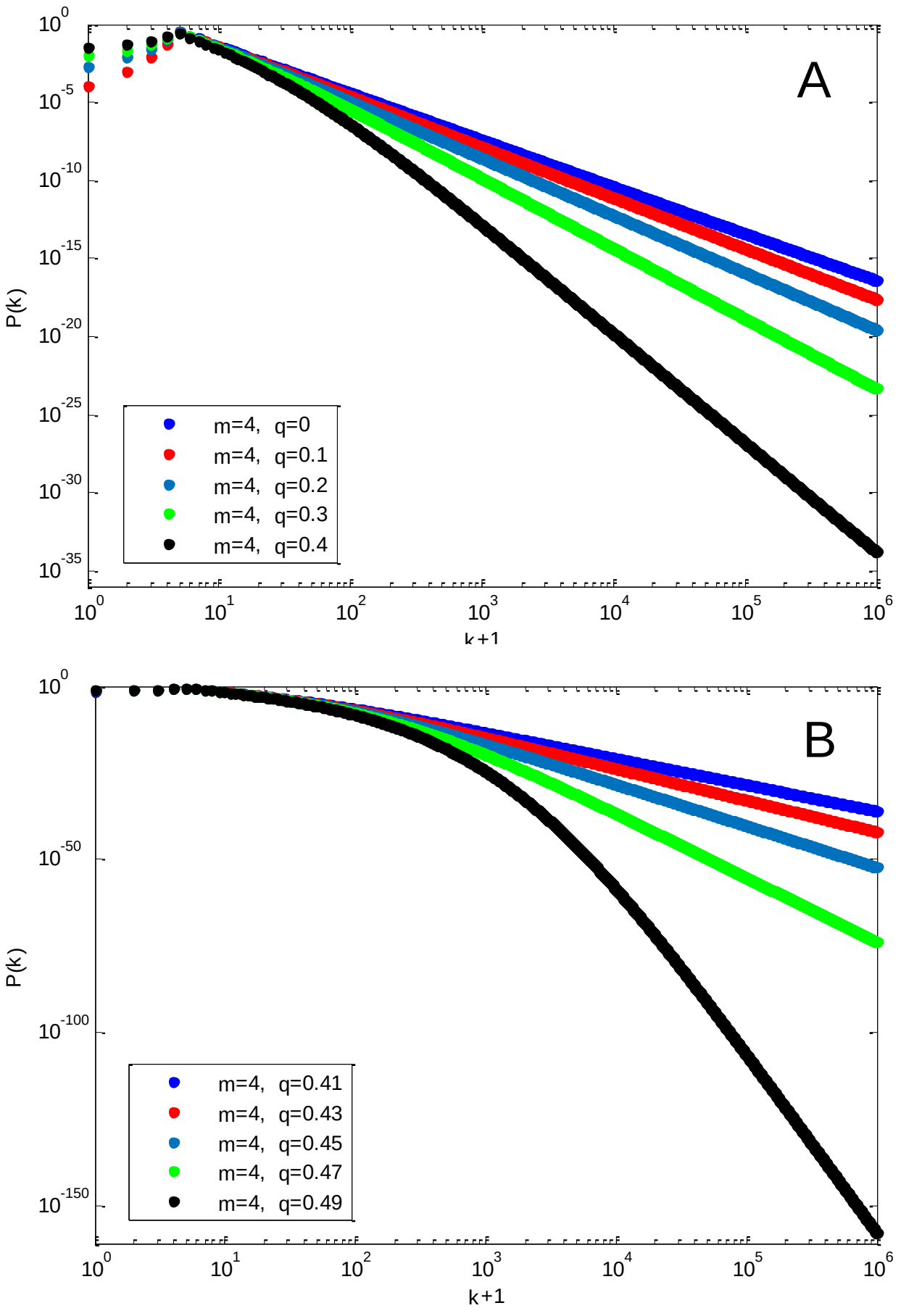

Fig. 1 The steady-state degree distribution for $m = 4$. (A) $0 \le q \le 0.4$ , (B) $0.4 < q < 0.5$

It is easy to find that when $k > 200$ (Fig. 1A) and $k > 3000$ (Fig. 1B), the tails approximate to straight lines, implying the degree distributions of the networks exhibit distinct power law tails. Similar results are observed with other $m$ and $q$. Therefore, the proposed birth-and-death network model shows the power law feature, providing theoretically robust support for the ubiquity of power law feature in real world networks.

Meanwhile, this study can also improve our understanding of power exponent $\alpha$ . According to our model, for sufficiently large $k$, we have $P(k) \propto k^{\alpha}$, where $\alpha$ can be obtained directly as follows (Appendix IV).

$$\alpha = -\frac{3-4q}{1-2q}$$

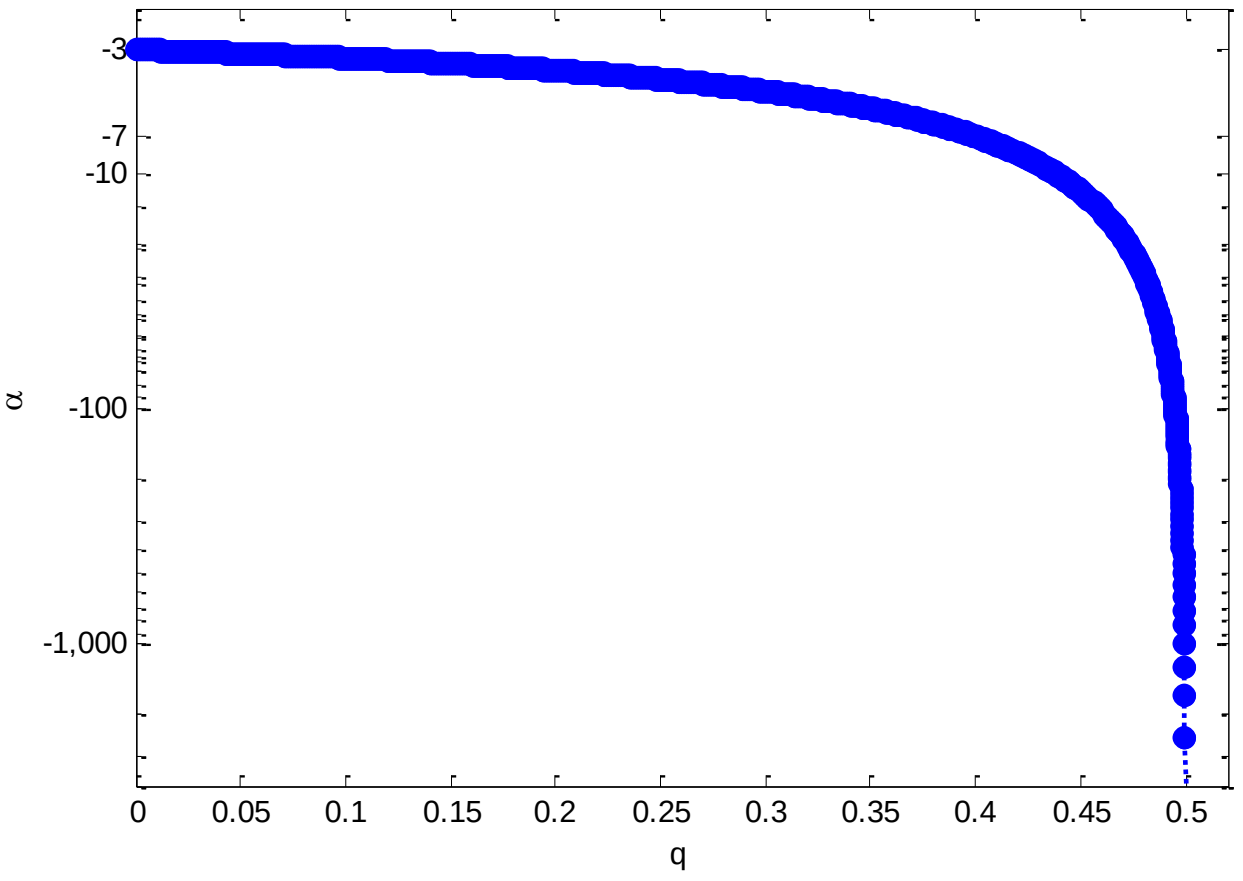

Fig.2 Relation between exponent $\alpha$ and node disappearance probability $q$

As illustrated in Fig.2, $\alpha$ will change with node disappearance probability $q$, explaining why various power law networks have different exponents in the real world. In particular, when $q = 0$, our evolving model degenerates into the BA model with $\alpha = -3$. Also we may find that $\alpha$ changes very slowly from −3 to −7 as $q$ is increasing from 0 to 0.4, but drops sharply once $q > 0.4$. In contrast with studies that highlight differing values of $\alpha$ resulting from linkage changes or aging (*7,8*), our findings emphasize the significant impact of $q$ on $\alpha$, and highlight their monotonic (decreasing) relationship.

***Why is the power law feature difficult to observe in real world networks?***

As we know, the power-law describes the statistical characteristics of nodes with large degrees. In general, since the probability of these nodes appearing in a complex network is relatively very small, a tiny sampling error may significantly affect their sampling frequencies, leading to a misjudgment about power law feature. Our theoretical results further show that as network size $n$ decreases or node disappearance probability $q$ increases, the power-law property become increasingly difficult to observe in the real world network.

For any empirical study, network data is critical for observing the power law feature of a network. Generally speaking, there are two types of data: whole network data and sampling data. In the case empirical data is the whole network data, network size $n$ will affect the deviation of its degree distribution tail from the power law tail. Indeed, the power law tail is obtained as $n \to +\infty$, meaning the smaller the network size $n$, the larger the deviation. As Fig. 3 shows, for $n = 100$ and 500, the tails of degree distributions deviate greatly from the power law. However, with the increase of $n$, the tails of degree distributions show an asymptotic behavior, that is, the range of $k$ subjected to the power law feature is gradually wider with the growth of $n$. From Fig.3, we find that as $n$ grows from 2,000 to 10,000, the range of $k$ which follows the power law property is also increasing from $10 \le k \le 100$ to $10 \le k \le 200$. Compared with the high degree cutoffs (*21,28*), this asymptotic behavior provides a dynamic lens for observing the power law characteristic of a real world network.

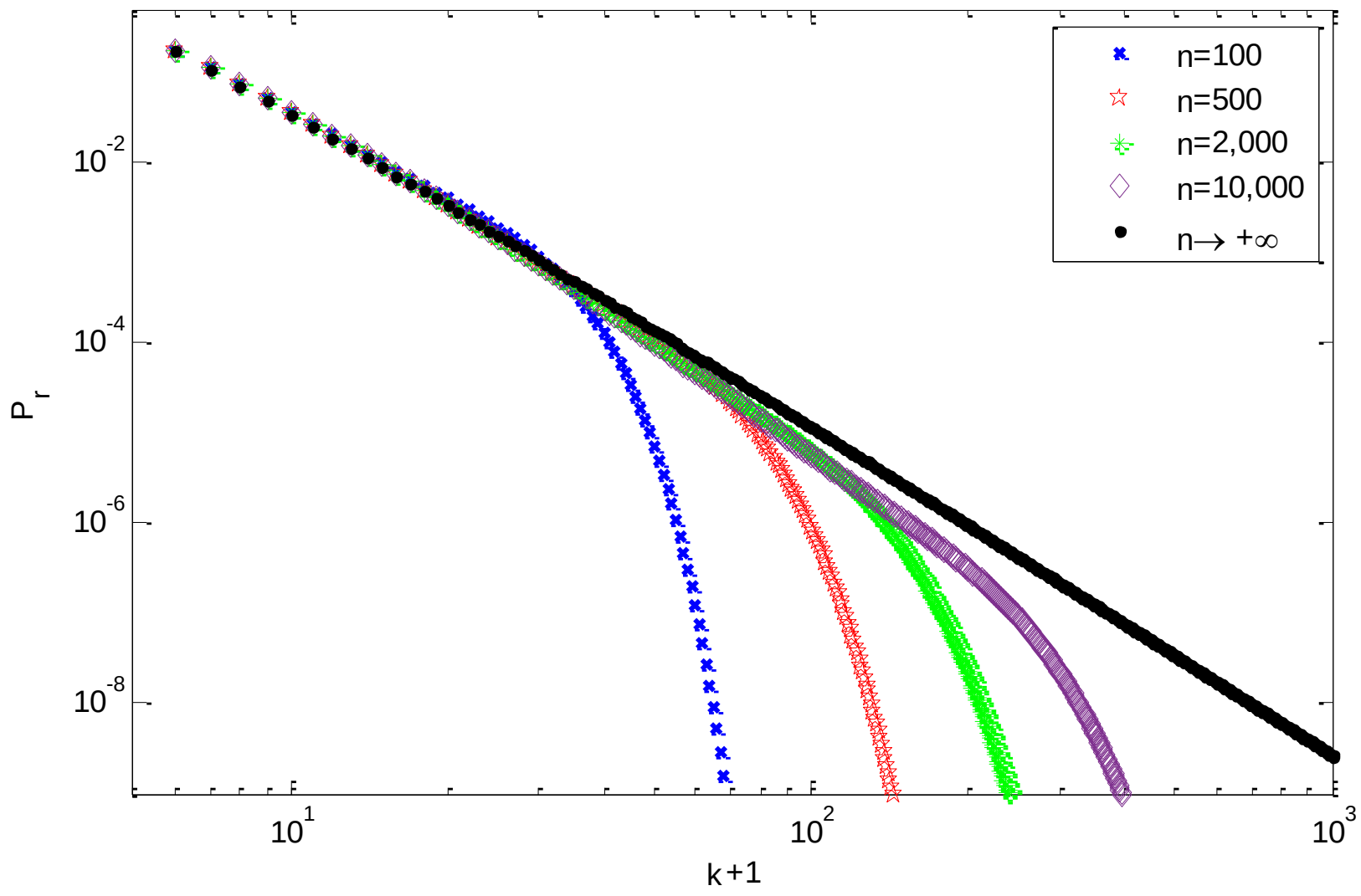

Fig. 3. The degree distributions for different network size $n$ under $m = 4$ and $q = 0.2$

In the case empirical data is obtained by sampling, in addition to the impact of network size $n$, the node disappearance probability $q$ will also affect the effective interval for detecting power law characteristics. Commonly in these empirical studies, frequency $f_k$ is used as a proxy for $P(k)$, and for any sampling data, there is a sampling error $\Delta$, satisfying $|f_k - P(k)| \le \Delta$, i.e. $0 \le f_k \le \Delta + P(k)$. As $\lim_{k \to +\infty} P(k) = 0$, for a fixed $q$, there exists a specific degree $k_q$, for any $k > k_q$, $P(k) \le 0.1\Delta$. Thus, for $k > k_q$, we have $0 \le f_k \le \Delta + P(k) \approx \Delta$, implying that $f_k$ cannot be employed as a proxy for $P(k)$. Therefore, the effective interval to observe the power law feature is $[m, k_q]$. Moreover, with the increase of $q$, $k_q$ is gradually decreasing, and the effective interval $[m, k_q]$ will be narrower, making more difficult the discernment of the power law feature.

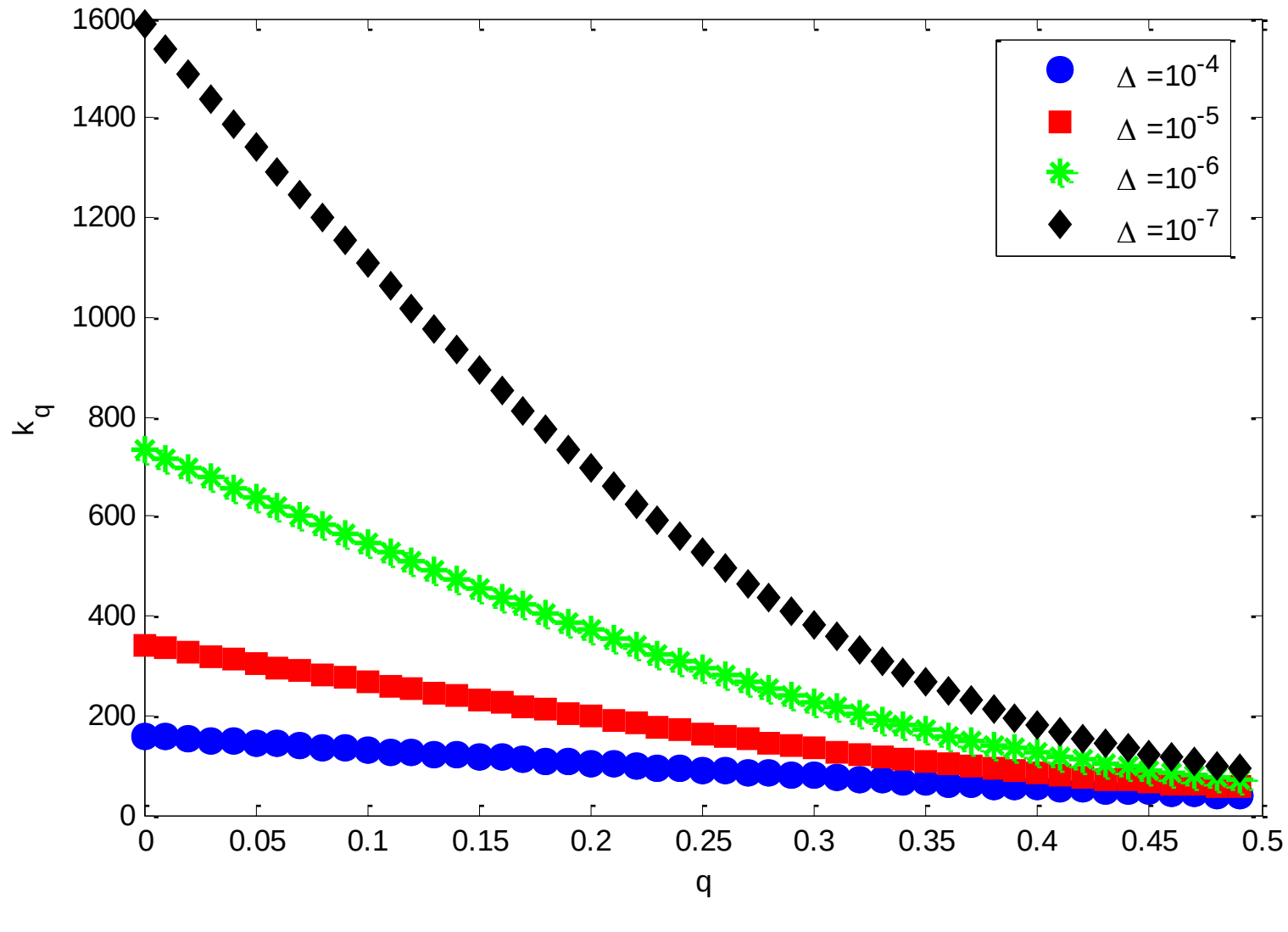

Fig. 4. Relationship between $k_q$ and $q$ for different $\Delta$ ($m$=4)

Fig. 4 shows the change of $k_q$ with $q$ under different sampling errors. Taking sampling error $\Delta = 10^{-7}$ (*1,9,14,27*) as an example, for $q = 0$, we have $k_q = 1587$ ,meaning the effective interval for detecting the power law feature is $[4,1587]$. Further, when $q \ge 0.4$, we have $k_q \le 182$, which contradicts the results of Fig. 1 that the power law tails are observed only for $k > 200$. Thus we suggest that for $q \geq 0.4$, the power law feature is imperceptible, i.e. almost impossible to perceive in an empirical study. Notice that for $q$=0.4, we have $\alpha = -7$, showing that $\alpha$ is between $-7$ and $-3$ for the observed networks. This theoretical result is consistent with empirical findings with $\alpha \in [-8, -2]$ in (*14-21*), which also shows the reliability and validity of our proposed model.

Combining the influences of both *n* and *q*, we suggest that a practical way to detect the power law property of a real world network is to observe the asymptotic behavior of the degree distribution within its effective interval $[m, k_q]$.

Thus, we conclude that the power law feature is proved to be universal in theory but difficult to observe in reality. Complex networks in the real world exhibit diverse evolving mechanisms (*45-50*). Although random addition and deletion of nodes and preferential attachment have been used to establish our birth-and-death network model, it is necessary to investigate other evolving rules and examine their limit properties *(48-50)*. Such further studies may help enrich our understanding of the power law mechanism as well as reveal more nuanced features.

# Supplementary Materials for

## The Truth about Power Law: Theory and Reality

**This supplementary file includes:**

Supplementary Text：Appendixes I~IV

## Appendix I: The Evolving Network Model with Preferential Attachment and Random Addition and Deletion of Nodes

### Model

The evolving rules of the birth-and-death network model characterized by preferential attachment and random addition and deletion of nodes are as follows:

(i) the initial network is a complete graph with $m+1\left(m\geq 1\right)$ nodes;

(ii) at each unit of time, randomly delete a node from the network with probability $q\left(0\leq q<1/2\right)$; or add a new node to the network with probability $p=1-q$ and connect it with $m$ old nodes of network by preferential attachment. The probability that the new node connects with old node $i$ depends on the degree $k_i$ of node $i$, i.e.

$$\pi_i = \frac{mk_i}{\sum_j k_j}$$

### Note

(a) Consider in the real world, any network size has its lower bound $n_0$. Here we assume $n_0 = 1$. This assumption will not affect the power law feature of the model.

(b) If at time $t$, a node is deleted, then all the edges incident to the removed node are also removed from the network, thus the degree of its neighbors decreases by one.

(c) If at time $t$, a new node is added to the network and the network size is less than $m$, then the new node is connected with all old nodes.

## Appendix II: The Steady-state Degree Distribution Equations

According to the stochastic process rules (SPR) method (*24*), we use $(n,k)$ to describe the state of node $v$, where $n$ is the number of nodes in the network that contains $v$, and $k$ is the degree of node $v$. Let $NK(t)$ be the state of node $v$ at time $t$. The stochastic process $\{NK(t), t \ge 0\}$ is an inhomogeneous Markov chain, and the state space is $E = \{(n,k), n \ge 1, 0 \le k \le n-1\}$. Let $\boldsymbol{P}$ be the one-step transition probability matrix of $\{NK(t), t \ge 0\}$ at time $t$,

$$\boldsymbol{P} = \left( p_{(n_1,k_1),(n_2,k_2)} \right) \tag{2.1}$$

Using SPR, the one-step transition probability matrix $\boldsymbol{P}$ produces two cases:

**1.** Add a node and link it to the old nodes by preferential attachment

(i) The one-step transition probability which $(n,k)$ turns to $(n+1,m)$ or $(n+1,n)$ is given by:

$$p_{(n,k),(n+1,m)} = P\left\{NK(t+1) = (n+1,m) \middle| NK(t) = (n,k)\right\} = \frac{p}{n+1}, \quad n \ge m+1, 0 \le k < n \tag{2.2}$$

$$p_{(n,k),(n+1,n)} = P\left\{NK(t+1) = (n+1,n) \middle| NK(t) = (n,k)\right\} = \frac{p}{n+1}, \quad n \le m, 0 \le k < n \tag{2.3}$$

(ii) The one-step transition probability which $(n,k)$ turns to $(n+1,k+1)$ is given by

$$\begin{aligned} p_{(n,k),(n+1,k+1)} &= P\left\{NK(t+1) = (n+1,k+1) \middle| NK(t) = (n,k)\right\} \\ &= \frac{mkp\sum_r P\{NK(t) = (n,r)\}}{(n+1)\sum_i iP\{NK(t) = (n,i)\}}, \quad n \ge m+1, 0 \le k < n \end{aligned} \tag{2.4}$$

$$p_{(n,k),(n+1,k+1)} = P\left\{NK(t+1) = (n+1,k+1) \middle| NK(t) = (n,k)\right\} = \frac{n}{n+1} p, \quad n \le m, 0 \le k < n \tag{2.5}$$

(iii) The one-step transition probability which $(n,k)$ turns to $(n+1,k)$ is given by

$$p_{(n,k),(n+1,k)} = P\left\{NK(t+1)=(n+1,k)\middle|NK(t)=(n,k)\right\}$$
$$= \frac{np}{n+1}\left[1-\frac{mk\sum_r P\left\{NK(t)=(n,r)\right\}}{n\sum_i iP\left\{NK(t)=(n,i)\right\}}\right], \quad n \geq m+1, 0 \leq k < n \tag{2.6}$$

**2.** Delete a node randomly

(iv) The one-step transition probability which $(n,k)$ turns to $(n-1,k-1)$ is given by

$$p_{(n,k),(n-1,k-1)} = P\left\{NK(t+1)=(n-1,k-1)\middle|NK(t)=(n,k)\right\} = \frac{k}{n-1}q, \quad 1 \leq k < n \tag{2.7}$$

$$p_{(1,0),(1,0)} = P\left\{NK(t+1)=(1,0)\middle|NK(t)=(1,0)\right\} = q \tag{2.8}$$

(v) The one-step transition probability which $(n,k)$ turns to $(n-1,k)$ is given by

$$p_{(n,k),(n-1,k)} = P\left\{NK(t+1)=(n-1,k)\middle|NK(t)=(n,k)\right\} = \frac{n-1-k}{n-1}q, \quad n-1 \geq k \geq 0, n \geq 2 \tag{2.9}$$

Let $\tilde{P}(t)=\left(p_{(n,k)}(t)\right)$ be the probability vectors of $NK(t)$ respectively, that is

$$p_{(n,k)}(t) = P\left\{NK(t)=(n,k)\right\} \tag{2.10}$$

the initial probability vector $\tilde{P}(0)=\left(p_{(n,k)}(0)\right)$ satisfies

$$p_{(m+1,m)}(0) = P\left\{NK(0)=(m+1,m)\right\} = 1 \tag{2.11}$$

We have

$$\tilde{P}(t+1) = \tilde{P}(t)\boldsymbol{P} \tag{2.12}$$

Let $K(t)$ be the average degree distribution at time *t*. Then we have

$$P\left\{K(t)=k\right\} = \sum_{i=k+1}^{+\infty} P\left\{NK(t)=(i,k)\right\} = \sum_{i=k+1}^{+\infty} \tilde{P}_{(i,k)}(t) \tag{2.13}$$

Thus the steady state degree distribution $K$ is given by

$$P(k) = \lim_{t\to+\infty} P\{K(t)=k\} = \lim_{t\to+\infty} \sum_{i=k+1}^{+\infty} P\{NK(t)=(i,k)\} \tag{2.14}$$

Noting in the case of $0 \le q < \frac{1}{2}$,

$$\lim_{n\to+\infty} \lim_{t\to+\infty} \sum_i P\{NK(t)=(n,i)\} = 1 \tag{2.15}$$

and

$$\lim_{n\to+\infty} \sum_i iP\{NK(t)=(n,i)\} = 2mp = 2m(1-q) \tag{2.16}$$

we can obtain the equations of steady-state degree distribution $K$ as follows

$$\begin{cases} pP(0) = qP(1) \\ \left(p + \frac{m}{2} + mq\right)P(m) = (m+1)qP(m+1) + \frac{m-1}{2}P(m-1) + p \\ \left(p + \frac{k}{2} + kq\right)P(k) = (k+1)qP(k+1) + \frac{k-1}{2}P(k-1) \qquad 1 \le k, k \ne m \end{cases} \tag{2.17}$$

When $q = 0$, the equations of steady-state degree distribution $K$ are as follows:

$$\begin{cases} (m+2)P(m) = 2 \\ (r+2)P(r) = (r-1)P(r-1) \qquad r \ge m+1 \end{cases} \tag{2.18}$$

Eq. (2.23) is also the equations of BA model.

## Appendix III: The Exact Solution of Steady-state Degree Distribution Equations

First we need to normalize equations (2.17) and then calculate the normalize equations using probability generating function. Here let

$$\Pi(k) = P(k) + \beta_k \qquad k = 0,1,2,\cdots$$

where $\beta_k = 0 \quad (k \geq m)$, and for $0 \leq k < m$, $\beta_k$ satisfies the following linear equations

$$\boldsymbol{A\beta} = \boldsymbol{\eta} \tag{3.1}$$

where

$$\boldsymbol{A} = \begin{bmatrix} -a_0 & b_1 & 0 & & & & \\ & c_1 & -a_2 & b_3 & & & \\ & & c_2 & -a_3 & b_4 & & \\ & & & \ddots & \ddots & \ddots & \\ & & & & c_{m-3} & -a_{m-2} & b_{m-1} \\ & & & & & c_{m-2} & -a_{m-1} \\ & & & & & & c_{m-1} \end{bmatrix}, \quad \boldsymbol{\beta} = \begin{bmatrix} \beta_0 \\ \beta_1 \\ \beta_2 \\ \vdots \\ \beta_{m-3} \\ \beta_{m-2} \\ \beta_{m-1} \end{bmatrix}, \quad \boldsymbol{\eta} = \begin{bmatrix} 0 \\ 0 \\ 0 \\ \vdots \\ 0 \\ 0 \\ p \end{bmatrix}_{m\times 1}$$

$$a_i = p + \frac{i}{2} + iq, b_i = iq, c_i = \frac{i}{2} \ \ i = 0,1,2,3,\cdots$$

So

$$\boldsymbol{\beta} = \boldsymbol{A}^{-1}\boldsymbol{\eta}$$

Then Eqs. (2.22) can be written as:

$$\begin{cases} p\Pi(0) = q\Pi(1) \\ \frac{3}{2}\Pi(1) = 2q\Pi(2) + \hat{p} \\ [2+q]\Pi(2) = 3q\Pi(3) + \frac{1}{2}\Pi(1) \\ \qquad\vdots \\ \left(p + \frac{r}{2} + rq\right)\Pi(r) = (r+1)q\Pi(r+1) + \frac{r-1}{2}\Pi(r-1) \\ \qquad\vdots \end{cases} \tag{3.2}$$

where $\hat{p} = \frac{3}{2}\beta_1 - 2q\beta_2$.

To solve Eqs. (3.2), we use the probability generation functions method. Let

$$G(x) = \sum_{k=0}^{\infty} \Pi(k)x^k \tag{3.3}$$

Multiplying $x^k$ to both sides of the ($k$+1)th equation in Eq. (3.2), and adding all of them respectively, we have

$$2pG(x) = G'(x)\left(x^2 - (1-2q)x + 2q\right) + 2\hat{p}x \tag{3.4}$$

Solving Eq. (3.4), we can get

$$G(x) = \left(\frac{1-x}{2q-x}\right)^{\frac{2p}{1-2q}} \int_x^{2q} \frac{2\hat{p}t}{(1-t)(2q-t)}\left(\frac{2q-t}{1-t}\right)^{\frac{2p}{1-2q}} dt \tag{3.5}$$

we may have

$$\begin{aligned} G(x) &= \frac{2\hat{p}q}{p} - 2\hat{p}\sum_{i=0}^{\infty} \frac{1}{\frac{2p}{1-2q} + i + 1}\left(\frac{2q-x}{1-x}\right)^{i+1} \\ &= \frac{2\hat{p}q}{p} - 2\hat{p}\sum_{i=0}^{\infty} \frac{1}{\frac{2p}{1-2q} + i + 1}\left(1 + \frac{2q-1}{1-x}\right)^{i+1} \end{aligned} \tag{3.6}$$

Employing Taylor expansion for Eq. (3.6), we get

$$G(x)=\frac{2\hat{p}q}{p}-2\hat{p}\sum_{i=0}^{+\infty}\frac{(2q)^{i+1}}{\frac{2p}{1-2q}+i+1}+$$

$$2\hat{p}\sum_{r=1}^{+\infty}\left[\sum_{i=1}^{+\infty}\frac{1}{\frac{2p}{1-2q}+i}\sum_{j=1}^{i}(-1)^{j+1}(1-2q)^{j}C_i^jC_{j+r-1}^r\right]x^r \tag{3.7}$$

Comparing with Eq. (3.3), $\Pi(k)$ can be obtained as follows:

$$\Pi(k)=\begin{cases}\dfrac{2\hat{p}q}{p}-2\hat{p}\displaystyle\sum_{i=0}^{+\infty}\frac{(2q)^{i+1}}{\frac{2p}{1-2q}+i+1} & k=0\\ 2\hat{p}\displaystyle\sum_{i=1}^{+\infty}\frac{1}{\frac{2p}{1-2q}+i}\sum_{j=1}^{i}(-1)^{j+1}(1-2q)^{j}C_i^jC_{j+k-1}^k & k\geq 1\end{cases} \tag{3.8}$$

Then the solution of Eqs. (2.17) is

$$P(k)=\begin{cases}\dfrac{2\hat{p}q}{p}-2\hat{p}\displaystyle\sum_{i=0}^{+\infty}\frac{(2q)^{i+1}}{\frac{2p}{1-2q}+i+1}-\beta_0 & k=0\\ 2\hat{p}\left[\displaystyle\sum_{i=1}^{+\infty}\frac{1}{\frac{2p}{1-2q}+i}\sum_{j=1}^{i}(-1)^{j+1}(1-2q)^{j}C_i^jC_{j+k-1}^k\right]-\beta_r & 1\leq k\leq m-1\\ 2\hat{p}\left[\displaystyle\sum_{i=1}^{+\infty}\frac{1}{\frac{2p}{1-2q}+i}\sum_{j=1}^{i}(-1)^{j+1}(1-2q)^{j}C_i^jC_{j+k-1}^k\right] & k\geq m\end{cases} \tag{3.9}$$

## Appendix IV: The Exponent of the Power Law

From Eqs. (2.17), for sufficiently large *k*, we have

$$\left(2p+k+2kq\right)P\left(k\right)=2\left(k+1\right)qP\left(k+1\right)+\left(k-1\right)P\left(k-1\right) \tag{4.1}$$

Noticing $P\left(k\right)\neq 0$, then

$$\frac{2q\left[kP\left(k\right)-\left(k+1\right)P\left(k+1\right)\right]}{P\left(k\right)}=\frac{\left(k-1\right)P\left(k-1\right)-kP\left(k\right)}{P\left(k\right)}-2p \tag{4.2}$$

Let

$$P\left(k\right)\propto \lambda k^{\alpha} \tag{4.3}$$

Taking (4.3) into (4.2), and let $k\to+\infty$, we obtain

$$\lim_{k\to+\infty}\frac{2q\left[k^{\alpha+1}-\left(k+1\right)^{\alpha+1}\right]}{k^{\alpha}}=\lim_{k\to+\infty}\frac{\left(k-1\right)^{\alpha+1}-k^{\alpha+1}}{k^{\alpha}}-2p \tag{4.4}$$

that is,

$$2q\left(\alpha+1\right)=\alpha+1+2p \tag{4.5}$$

Hence

$$\alpha=-\frac{3-4q}{1-2q} \tag{4.6}$$

Since $\alpha$ is monotonically decreasing with $q$, it is easy to find that $\alpha\leq-3$ for all $0\leq q<\frac{1}{2}$. Specially, if and only if $q=0$, we have $\alpha=-3$.